\documentclass[twocolumn,pre,showpacs,amssymb,nofootinbib]{revtex4}
\usepackage{graphics,graphicx}

\begin{document}
\title{Multiple Choice Minority Game With Different Publicly Known Histories}
\author{H.~F. Chau}
\email{hfchau@hkusua.hku.hk}
\author{F.~K. Chow}
\author{K.~H. Ho}
\author{W.~C. Man}
\affiliation{Department of Physics, University of Hong Kong, Pokfulam Road,
 Hong Kong}
\affiliation{Center of Theoretical and Computational Physics, University of
 Hong Kong, Pokfulam Road, Hong Kong}
\date{\today}

\begin{abstract}
 In the standard Minority Game, players use historical minority choices as the
 sole public information to pick one out of the two alternatives. However,
 publishing historical minority choices is not the only way to present global
 system information to players when more than two alternatives are available.
 Thus, it is instructive to study the dynamics and cooperative behaviors of
 this extended game as a function of the global information provided. We
 numerically find that although the system dynamics depends on the kind of
 public information given to the players, the degree of cooperation follows the
 same trend as that of the standard Minority Game. We also explain most of our
 findings by the crowd-anticrowd theory.
\end{abstract}

\maketitle
\section{Introduction}
 Many phenomena in a variety of fields including biology and economics can be
 modeled by agent-based complex adaptive systems (CAS) \cite{CAS1,CAS2,CAS3}.
 In fact, CAS can be used to gain empirical understanding, normative
 understanding, qualitative insight and theory generation together with
 methodological advancement in economic systems. This agent-based approach
 focuses on the dynamics and the effects of the initial or boundary conditions
 on an economic system as opposed to the conventional economic methodology
 which concentrates mainly on the equilibrium state of the system \cite{CAS3}.
 In this respect, techniques in statistical physics and non-linear dynamics can
 be applied in the study of economic systems. This is the goal of the rapidly
 growing field of econophysics.

 Minority Game (MG) \cite{MG} is perhaps the simplest agent-based econophysical
 model that captures the minority seeking behavior of independent selfish
 players. In the original version of MG, each player picks one out of two
 alternatives in each time step based on the publicly posted minority choices
 of the previous $M$ turns. Those correctly picking the minority choice are
 awarded one dollar while the others are deducted one dollar. Although players
 in MG are selfish and only aim at maximizing their own wealth, they do so in a
 cooperative manner. In addition, MG exhibits a second order phase transition
 point dividing the parameter space into the so-called symmetric and asymmetric
 phases \cite{MG,MG1}. Besides, the cooperation phenomenon and the phase
 transition point appear to be very common as they are also observed in several
 variants of MG that use more than two alternatives \cite{NcMG,MCMG}, evolving
 strategies \cite{Evo1,Evo2}, different payoff functions
 \cite{Payoff1,Payoff2}, different network topology \cite{NMG} and a mixed
 population of players \cite{Heter1,Heter2}.

 In MG, some public information is given to players for reference in making
 their decisions. In both the original MG and many of its variants, the public
 information is the minority choices of the previous $M$ turns. In other words,
 this public information gives a complete description of historical winning
 choices of the previous $M$ turns. Of course one may give other public
 information to players. Perhaps the most well-known case comes from a series
 of studies concerning the relevance of history in MG that was initiated by a
 paper of Cavagna \cite{Cavagna}. This series of studies investigated the
 effect of replacing the actual historical winning choices by some fake ones on
 the dynamics of MG. For instance, Challet and Marsili extensively studied the
 effect of substituting a randomly and independently chosen bit string for each
 historical minority choice on the dynamics of MG. They found that although the
 modified game still shows phase transition right at the same point in the
 parameter space, the dynamics is markedly different from the original MG in
 the asymmetric phase \cite{Memory}. Their findings were echoed by a recent
 work of Ho \emph{et al.} who discovered that the dynamics of the original and
 the modified game also differ in the symmetric phase \cite{LatestMemory}.
 Other examples of using different public information came indirectly from the
 investigations of players with different memory sizes
 \cite{MG,EnhancedWinning} and players acting only on local information
 \cite{NMG,LMG}. These studies showed that the dynamics in many variants of MG
 depends on the historical outcomes of the game.

 Nonetheless, publishing the historical winning choices of the previous $M$
 turns is not the only way to present certain real global information of the
 system to players. For example, it is instructive to investigate what will
 happen if the publicly known historical minority choice is replaced by the
 publicly known historical majority choice. In the case of the original MG
 \cite{MG,MG1}, the statistical properties of this majority history model is
 identical to that of the original MG as the knowledge of the historical
 majority choice is equivalent to that of the historical minority choice. In
 contrast, the situation is radically different when the number of alternatives
 in the model is greater than two. An extension of MG that allows more than two
 alternatives was proposed by Ein-Dor \emph{et al.} \cite{EinDor}. They
 regarded the alternatives as states of Potts spins and replaced player's
 strategies by feedforward neural networks with Hebbian learning rules. Their
 model is not suitable to study the effect of global information replacement as
 the method used by players to decide their choices is also changed. Thus, our
 study is based on a simpler extension of the standard MG proposed by Chau and
 Chow \cite{NcMG} that allows players to choose from $N_c > 2$ equally capable
 alternatives with different types of publicly posted real histories without
 changing the algorithm of player's decision. Our numerical simulations show
 that the general trend of the cooperative behavior does not depend on the kind
 of common real historical data used. In addition, we find that the location of
 the second order phase transition point separating the symmetric and
 asymmetric phases is independent of the publicly posted histories. Most of our
 findings can be understood by the crowd-anticrowd theory proposed by Hart
 \emph{et al.} \cite{Crowd1,Crowd2}.

\section{The Game MG$(N_c)$ And Its Extension}
 Our study focuses on an extension of the MG known as MG$(N_c)$ proposed by
 Chau and Chow in Ref.~\cite{NcMG}. In MG$(N_c)$, each of the $N$ players picks
 one out of $N_c$ alternatives independently in each turn where $N_c$ is a
 prime power.\footnote{There are two reasons why we restrict $N_c$ be to a
 prime power. (Definitions of various terms in this footnote can be found later
 in the text.) First, finding the maximal reduced strategy space size is a very
 difficult combinatorial problem whose solution is not known for a general
 $N_c$ to date. Second, even if the maximal reduced strategy space is found, it
 is possible that such space is not uniformly distributed in the full strategy
 space. Hence, the expression of the control parameter $\alpha$ is not known
 when $N_c$ is not a prime power.} The choice picked by the least
 \emph{non-zero} number of players is said to be the (1st) minority choice in
 that turn. Those who have picked the minority choice will be awarded one
 dollar while the others will be deducted one dollar. (In this respect, the
 $N_c$ alternatives are treated on equal footing in this game.) The minority
 choices of the previous $M$ turns are publicly announced. To aid each player
 making their decisions, each of them is randomly and independently assigned
 once and for all $S$ deterministic strategies before the game begins. A
 strategy is a table that assigns every possible history (in this case, the 1st
 minority choice of the previous $M$ turns) to a choice. In other words, it is
 a map from the set of all possible histories $H$ to the finite field of $N_c$
 elements $GF(N_c)$. Clearly, there are totally $N_c^{N_c^M}$ different
 possible strategies and this collection of strategies is called the full
 strategy space. To evaluate the performance of a strategy, a player looks at
 the virtual score which is the current hypothetical wealth if that strategy
 were used throughout the game. Every player follows the choice of his/her
 current best working strategy, namely, the one with the largest virtual score,
 to pick an alternative \cite{NcMG}. (In case of a tie, the player randomly
 picks one from his/her pool of best working strategies.)

 Now, we consider a generalization of the MG$(N_c)$ model known as
 MG$_\textrm{\scriptsize sub}(N_c)$, where the subscript
 ``$\textrm{sub}$'' describes the kind of historical choices used. When
 $\textrm{sub} = \min (q)$, we publicly announce the historical $q$th minority
 choices of the past $M$ turns instead of the historical 1st minority choices.
 (More precisely, we arrange those alternatives chosen by non-zero number of
 players in ascending order of the number of players chosen. Those alternatives
 with equal number of players chosen are arranged randomly in this sequence.
 The $q$th minority choice is the $q$th alternative in this sequence. In the
 event that the number of alternatives chosen by non-zero number of players is
 less than $q$, we define the $q$th minority choice as the last entry in this
 sequence.) Similarly, when $\textrm{sub} = \textrm{maj}(q)$, we publish the
 historical $q$th majority choices of the past $M$ turns. We call the publicly
 announced alternatives the history string irrespective of the state ``sub''.
 Moreover, we stress that apart from the global information released, all the
 rules in MG$_\textrm{\scriptsize sub}(N_c)$ are the same as those of
 MG$(N_c)$. Thus, MG$(N_c) = $MG$_{\min (1)}(N_c)$.

\section{Numerical Results}
 Following Refs.~\cite{MG,NcMG}, we measure the degree of player cooperation by
 considering the mean variance of attendance over all alternatives (or simply
 the mean variance)
\begin{eqnarray}
 \Sigma^2 & = & \frac{1}{N_c} \sum_{i\in GF(N_c)} \left< \left[ \left<
  (A_i(t)^2 \right>_t - \left< A_i(t) \right>^2_t \right] \right>_\Xi \nonumber
  \\
 & = & \left< \left[ \left< (A_0(t))^2 \right>_t - \left< A_0(t)
  \right>^2_t \right] \right>_\Xi , \label{E:variance_Def}
\end{eqnarray}
 where the attendance of an alternative $A_i(t)$ is the number of players
 picking the alternative $i$ at turn $t$. Note that $\left< \cdots \right>_t$
 and $\left< \cdots \right>_\Xi$ are the expectation values averaged over time
 and over strategies initially assigned to the players respectively. Since all
 the $N_c$ alternatives are treated on equal footing in
 MG$_\textrm{\scriptsize sub}(N_c)$, we may arrived at the second line in
 Eq.~(\ref{E:variance_Def}). The smaller the $\Sigma^2$, the better the player
 cooperates. More importantly, for a fixed $S$ and up to first order
 approximation, $\Sigma^2$ depends only on the control parameter $\alpha =
 N_c^{M+1} / NS$ which measures the relative diversity of the strategies used
 in the system \cite{Savit}.

 Furthermore, to investigate the phase diagram of
 MG$_\textrm{\scriptsize sub}(N_c)$, we follow Refs.~\cite{MCMG,Order} to study
 the order parameter
\begin{equation}
 \theta = \frac{1}{N_c^M} \sum_\mu \left\{ \sum_\Omega \left[ \left< p(\Omega |
 \mu) \right>_t - \frac{1}{N_c} \right]^2 \right\} , \label{E:order_Def}
\end{equation}
 where $\left< p(\Omega | \mu) \right>_t$ denotes the time average of the
 probability that the current minority choice is $\Omega$ conditioned on a
 global history string $\mu$.

\begin{figure}[t]
 \begin{center}
  \includegraphics*[scale=0.31]{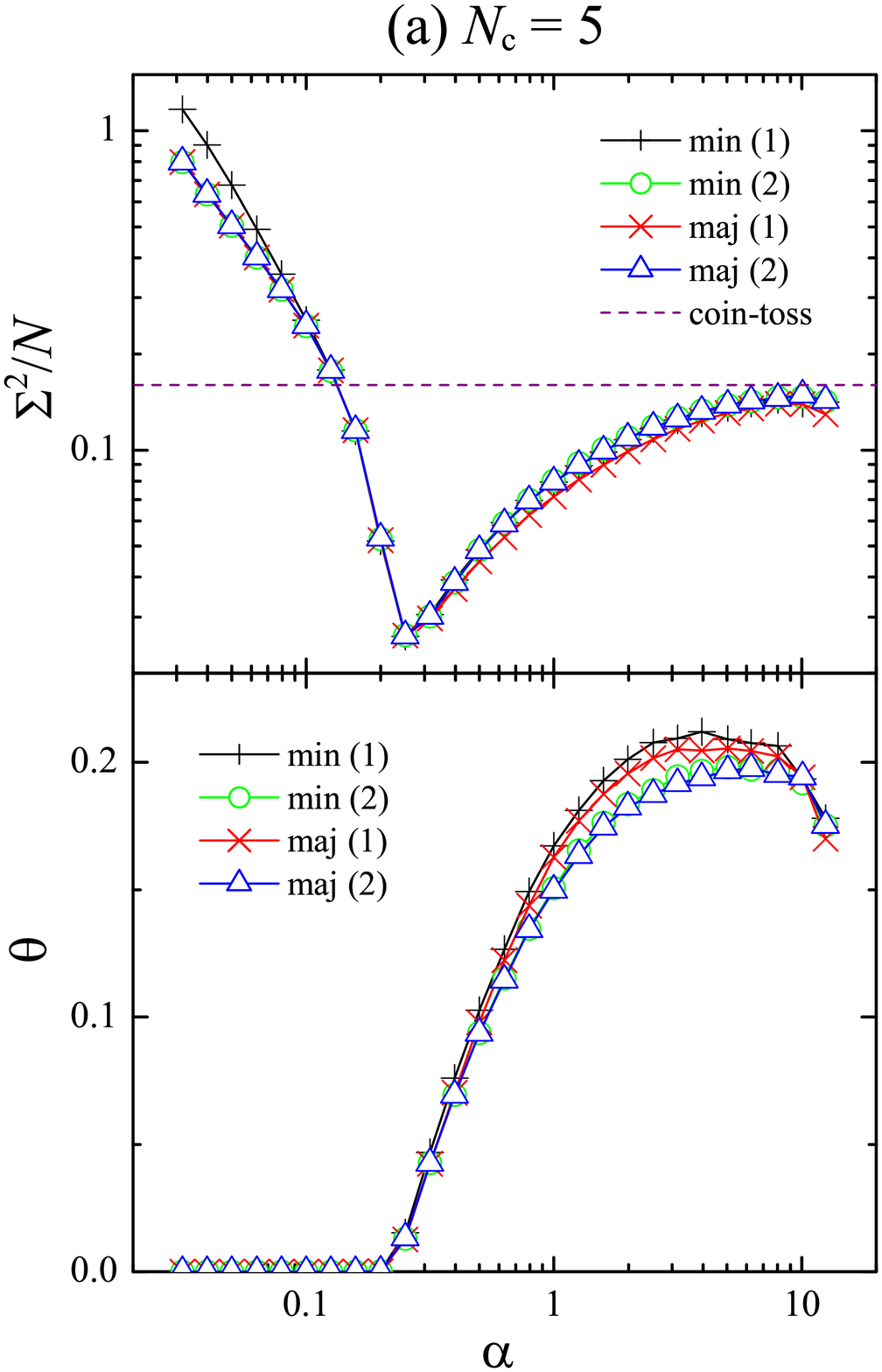}
  \includegraphics*[scale=0.31]{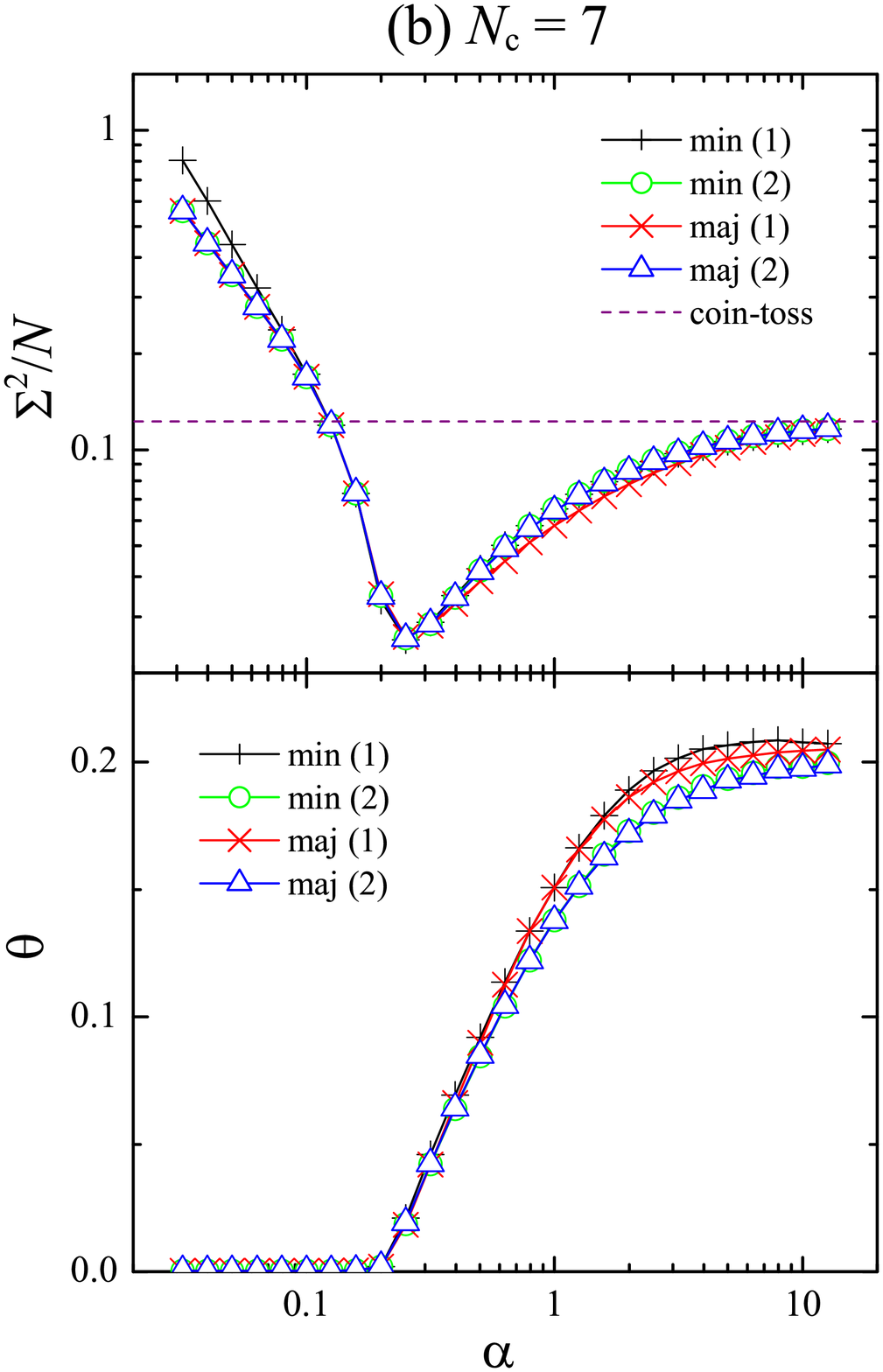}
 \end{center}
 \caption{The variance of attendance per player $\Sigma^2/N$ and the order
  parameter $\theta$ \emph{vs.} the control parameter $\alpha$ for
  (a)~MG$_\textrm{\scriptsize ``sub''}(5)$ and
  (b)~MG$_\textrm{\scriptsize ``sub''}(7)$. The values of $M$ and $S$ used in
  all the figures in this paper are $3$ and $2$, respectively.}
 \label{F:Var}
\end{figure}

 Fig.~\ref{F:Var} plots $\Sigma^2 / N$ and $\theta$ as a function of $\alpha$
 for different ``$\textrm{sub}$'' and $N_c$. Each data point presents the
 average value over 1000 different runs and the value for each run is averaged
 over 25000 iterations after discarding the first 20000 iterations to allow
 equilibration. We find that within the range of parameters we have simulated,
 the system indeed equilibrates well before the first 20000 iterations. All
 $\Sigma^2 / N$ curves in Fig.~\ref{F:Var} show a similar trend and have cusps
 around $\alpha \approx 1$ irrespective of the values of $N_c$ and ``sub''
 used; and the $\theta$ curves indicate second order phase transitions around
 the cusps separating the symmetric and the asymmetric phases. (The small drop
 in $\Sigma^2 / N$ for large value of $\alpha$ in Fig.~\ref{F:Var}a is due to
 finite size effect as the number of players $N$ is about 30.) Although we only
 show those curves for $N_c = 5$ and $7$, similar behaviors are observed for
 other values of $N_c > 2$. In fact, our numerical simulations show that the
 critical points of all curves with the same $N_c$ and $S$ coincide. That is to
 say, the critical value $\alpha_c$ is found to be a function of $N_c$ and $S$
 only and is independent of the kind of history string used. Besides, $\left.
 \Sigma^2 / N \right|_{\alpha_c}$ is a function of $N_c$ and $S$ only. The
 behavior of $\Sigma^2 / N$ away from the phase transition point $\alpha_c$ is
 also worth mentioning. For $\alpha \ll \alpha_c$, the variance per player for
 $\textrm{sub} = \min (1)$ is consistently greater than those obtained in MGs
 using other public information. In contrast, for $\alpha \gtrsim \alpha_c$,
 $\Sigma^2 / N$ for $\textrm{sub} = \textrm{maj} (1)$ is consistently smaller
 than those obtained in MGs using other globally announced information.

\section{The Possibility Of An Analytical Or Semi-Analytical Solution}
 Three major approaches to study the MG analytically or semi-analytically are
 known to date. We briefly discuss their potentials in solving the
 MG$_\textrm{\scriptsize sub}(N_c)$ one by one below.

\subsection{Replica Trick}
 Replica trick was used by Challet \emph{et al.} to solve the MG analytically
 \cite{Order,Replica1,Replica2}. By using fact that knowing the majority choice
 automatically implies the knowledge of the minority choice when $N_c = 2$,
 they wrote down a simple Hamiltonian quadratic in the random spin variable.
 Standard replica trick can then be used to compute the probability
 distribution of the average action of a player. Subsequently, the quantities
 such as variance per player $\Sigma^2 / N$ can be computed analytically.
 Although the value of $\Sigma^2 / N$ obtained by replica trick agrees
 reasonably well with the numerical findings in the asymmetric phase of the
 standard MG, the averaging procedure in replica trick makes this method
 impossible to recover the dynamics and the value of $\Sigma^2 / N$ in the
 symmetric phase.

 Replica trick faces a further problem when $N_c > 2$. The Hamiltonian involves
 a term in the form $\delta_{i,i_0}$ for some $i,i_0 \in GF(N_c)$. This term is
 equal to the degree $(N_c -1)$ polynomial $\prod_{j\neq i_0} (i-j) / (i-i_0)$
 in random spin variable and hence the associated saddle point equation is much
 harder to solve.

\subsection{Generating Functional Method}
 Coolen pioneered the use of generating functional method to solve the standard
 MG \cite{BatchCoolen,Coolen}. This method is mathematically rigorous and
 exact. In fact, it can be used to compute the variance per player $\Sigma^2 /
 N$ in the asymmetric phase of MG efficiently. Nonetheless, the presence of
 long time scale periodic dynamics in the symmetric phase makes the computation
 of $\Sigma^2 / N$ in this phase using generation functional method not very
 successfully to date \cite{CoolenReply}.

 Applying generation functional method to solve
 MG$_\textrm{\scriptsize sub}(N_c)$ could face two more problems. First, the
 equation governing the evolution from one turn to the next involves a term in
 the form $\delta_{i,i_0}$ for some spin variables $i,i_0\in GF(N_c)$ and hence
 is equal to a polynomial of degree $(N_c - 1)$. Second, computing the relative
 popularity of the $N_c$ alternatives are required when $\textrm{sub} \neq
 \min(1)$ or $\textrm{maj}(1)$. These two requirements add further complexity
 to the generating functional when $N_c > 2$.

\subsection{Crowd-Anticrowd Theory}
 Developed by Hart \emph{et al.} \cite{Crowd1,Crowd2}, crowd-anticrowd theory
 is a semi-analytical method to explain the dynamics and variance observed in
 various variants of MG. This method provides an intuitive understanding of the
 origin of player cooperation. Although it predicts the existence of phase
 transition around $\alpha \approx 1$, it does not give us a simple way to
 calculate the value of $\alpha_c$.

 Chau \emph{et al.} \cite{NcMG,MCMG} have successfully extended the
 crowd-anticrowd theory to study the case when the number of alternatives $N_c$
 is a prime power. This makes the theory a good choice to investigate the
 dynamics of MG$_\textrm{\scriptsize sub}(N_c)$.

\section{The Crowd-Anticrowd Explanation} 
 Most of our numerical simulation results can be explained by the
 crowd-anticrowd theory. Recall that two strategies are said to be uncorrelated
 if the probability for them to make the same choice equals $1/N_c$ when
 averaged over the set of all possible history string. And two strategies are
 called anti-correlated if they make different choices for every input history
 string. Besides, two strategies are said to be significantly different if they
 are either anti-correlated or uncorrelated. In fact, one can form a subset of
 $N_c^{M+1}$ strategies from the full strategy space in such a way that any two
 distinct strategies in this subset are significantly different. Besides, the
 size of this subset is maximal in the sense that no such subset with more than
 $N_c^{M+1}$ strategies exists. This subset is called the maximal reduced
 strategy space \cite{MG1,NcMG,Crowd1,Crowd2}. Most importantly, numerical
 simulations show that the dynamics of MG$_\textrm{\scriptsize ``sub''}(N_c)$
 for strategies taken from the full or from the maximal reduced strategy spaces
 are similar.

 From the discussions in Ref.~\cite{NcMG}, one may label a strategy in the
 maximal reduced strategy space by $(\lambda,\beta) \in GF(N_c^M) \times
 GF(N_c)$ in such a way that two strategies $(\lambda,\beta)$ and $(\lambda',
 \beta')$ are uncorrelated if and only if $\lambda\neq\lambda'$. They are
 anti-correlated if and only if $\lambda=\lambda'$ and $\beta\neq \beta'$. They
 are the same if and only if $\lambda=\lambda'$ and $\beta= \beta'$. According
 to the crowd-anticrowd theory, the mean variance of attendance $\Sigma^2$ in
 MG$(N_c)$ is governed by an ensemble of mutually uncorrelated sets of
 anti-correlated strategies \cite{NcMG,Crowd1,Crowd2}. That is to say,
\begin{equation}
 \Sigma^2 \approx \left< \frac{1}{N_c^{M+2}} \sum_{\lambda,\beta} \left\{
 \sum_{\beta'\neq\beta} \left[ N_{\lambda,\beta}(t) - N_{\lambda,\beta'}(t)
 \right] \right\}^2 \right>_{t,\Xi} , \label{E:Sigma_CAC}
\end{equation}
 where $N_{\lambda,\beta}(t)$ denotes the number of players making decision
 according to the strategy $(\lambda,\beta)$ in the anti-correlated strategy
 set $\lambda$ in turn $t$.

\begin{figure}[t]
 \begin{center}
 \includegraphics*[scale=0.31]{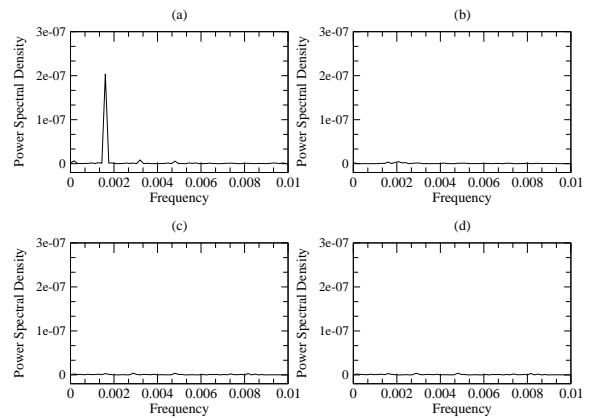}
 \end{center}
 \caption{The power spectral density of the auto-correlation function of
  the attendance $A(t)$ against frequency for a typical run in (a)~MG$_{\min
  (1)}(5)$, (b)~MG$_{\min (2)}(5)$, (c)~MG$_{\textrm{\scriptsize maj} (1)}(5)$
  and (d)~MG$_{\textrm{\scriptsize maj} (2)}(5)$ averaged by 50 runs for
  $\alpha = 0.05$. The period-$N_c^{M+1}$ dynamics is pronounced only in (a).}
 \label{F:Period_Dyn}
\end{figure}

 When $\alpha\ll \alpha_c$, known as the symmetric phase, there is a periodic
 dynamics in the time series of the minority choice. According to the
 crowd-anticrowd theory, this dynamics leads to a large mean variance of
 attendance per player $\Sigma^2 / N$ in MG$_{\min (1)}(N_c)$
 \cite{Crowd1,Crowd2,Savit,Manuca,Dynamics}. Let us briefly review the origin
 of this periodic dynamics in MG$_{\min (1)}(N_c)$. When the number of
 strategies at play is much larger than the maximal reduced strategy space size
 $N_c^{M+1}$, it is very likely for players to employ similar strategies.
 Initially, for a given history string $\mu$, every alternative has equal
 probability of being the minority. Moreover, the virtual score of a strategy
 $(\lambda,\beta)$ that gives the correct prediction of the minority is
 increased while that of its anti-correlated strategies are decreased. As there
 are more players than the maximal reduced strategy space size, in the next
 occurrence of the same history string $\mu$, more players may use $(\lambda,
 \beta)$ to pick their alternatives. Thus, $(\lambda,\beta)$ is less likely to
 correctly predict the minority choice due to overcrowding of strategies.
 Inductively, overcrowding of strategies leads to the existence of a
 period-$N_c$ dynamics in the minority choice as well as the attendance time
 series conditioned on an arbitrary but fixed history in MG$_{\min (1)}(N_c)$ 
 \cite{Crowd1,Crowd2,Memory,Dynamics,Manuca,LatestMemory}. Another periodic
 dynamics coming from a slightly different origin is also present in MG$_{\min
 (1)}(N_c)$. Recall that the history string gives complete information of the
 winning choices in the past $M$ turns in MG$_{\min (1)}(N_c)$ making its
 minority choice time series highly correlated. By extending the analysis of
 Challet and Marsili in Ref.~\cite{Memory} from $N_c = 2$ to a general prime
 power $N_c$, we conclude that the minority choice time series from the
 $(N_c^{M+1} k +1)$th to the $[N_c^{M+1} (k+1)]$th turn is likely to form a
 de~Bruijn sequence\footnote{A de~Bruijn sequence of length $q^n$ over an
 alphabet of size $q$ is defined as a sequence that contains all the $q^n$
 possible $n$-tuples as its subsequence. (We allow wraparound when defining a
 subsequence. For example, $0110$ is a de~Bruijn sequence of length 4 over the
 set of alphabets $\{ 0,1\}$.)} of length $N_c^{M+1}$ for all $k \in
 {\mathbb N}$ resulting in a period $N_c^{M+1}$ peak in the Fourier transform
 of both the minority choice and the attendance time series \cite{Dynamics}.
 Besides, it is likely that between the above $N_c^{M+1}$ turns, each strategy
 wins exactly $N_c^M$ times. We follow the convention in
 Ref.~\cite{LatestMemory} by calling this correlation in the minority choice
 and attendance time series the period-$N_c^{M+1}$ dynamics. Note that because
 of the period-$N_c^{M+1}$ dynamics, the virtual score difference between any
 two strategies is likely to be zero in the $(N_c^{M+1} k + 1)$th turn for all
 $k\in {\mathbb N}$. We call this phenomenon virtual score reset
 \cite{LatestMemory}.

\begin{figure}[t]
 \begin{center}
 \includegraphics*[scale=0.31]{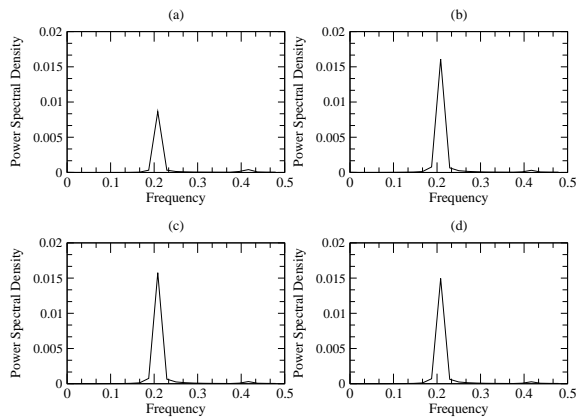}
 \end{center}
 \caption{The power spectral density of the auto-correlation function of
  the attendance $A(t)$ conditioned on an arbitrary but fixed history against
  frequency shows the slight strengthening of period-$N_c$ dynamics when $N_c
  > 2$. Parameters used in this plot is the same as that in
  Fig.~\ref{F:Period_Dyn}.}
 \label{F:Cond_Period_Dyn}
\end{figure}

\begin{figure}[t]
 \begin{center}
 \includegraphics*[scale=0.31]{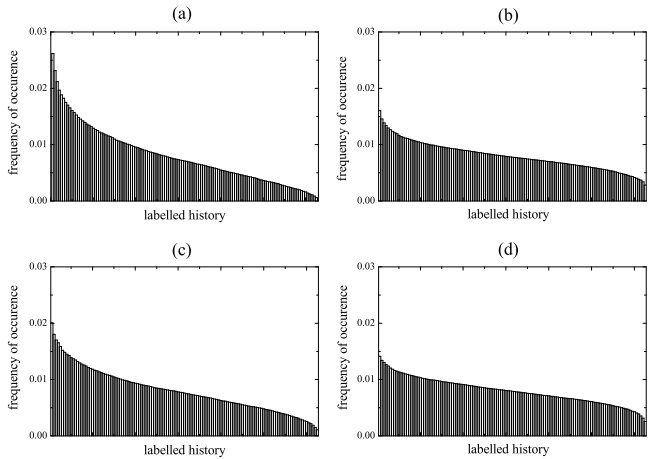}
 \end{center}
 \caption{We plot the frequency of occurrence of history strings sorted in
  descending order and then averaged over 50 runs for
  (a)~MG$_{\textrm{\scriptsize maj} (1)}(5)$, (b)~MG$_{\textrm{\scriptsize maj}
  (2)}(5)$, (c)~MG$_{\min (1)}(5)$ and (d)~MG$_{\min (2)}(5)$ at $\alpha =
  1.5$.}
 \label{F:Freq_Dist}
\end{figure}

 However, for MG$_\textrm{\scriptsize sub}(N_c)$ other than MG$_{\min
 (1)}(N_c)$ or MG$_\textrm{\scriptsize maj (1)}(2)$, the knowledge of the
 history string does not give a player complete information on the minority
 choice. Suppose again that initially the strategy $(\lambda,\beta)$ correctly
 predicts the minority choice for a given history string $\mu$. Since the
 virtual score calculation is still based only on the historical minority
 choices, in the next occurrence of $\mu$, strategy $(\lambda,\beta)$ may well
 be able to correctly predict the minority choice as it is chosen by only a few
 players. Therefore, the publicly announced histories from the $(N_c^{M+1} k +
 1)$th to the $[N_c^{M+1} (k+1)]$th turn no longer tend to form a de~Bruijn
 sequence and the virtual score difference between two distinct strategies is
 unlikely to reset \cite{Dynamics,LatestMemory}. That is why the
 period-$N_c^{M+1}$ dynamics almost completely disappears as shown in
 Fig.~\ref{F:Period_Dyn}. (Although we only present the periodic dynamics and
 distribution of history strings for the case of $N_c = 5$ in
 Figs.~\ref{F:Period_Dyn}--\ref{F:Freq_Dist}, our simulations for other values
 of $N_c$ are consistent with our crowd-anticrowd explanation in this section.)
 Nonetheless, from Fig.~\ref{F:Cond_Period_Dyn}, we observe that the
 period-$N_c$ dynamics is slightly strengthened. To understand why, let us
 recall that in MG$_{\min (1)}(N_c)$ and MG$_\textrm{\scriptsize maj (1)}(2)$,
 the virtual score reset mechanism implies that the $(N_c k + 1)$th terms for
 all $k\in {\mathbb N}$ in the attendance time series conditioned on an
 individual history is positively correlated. This is the major contributor to
 the period-$N_c$ dynamics. In contrast, for other
 MG$_\textrm{\scriptsize sub}(N_c)$ and for a fixed $\ell = 1,2,\ldots ,N_c$,
 the absence of a virtual score reset mechanism implies that correlations among
 the $(N_c k + \ell)$th terms for all $k\in {\mathbb N}$ in the time series of
 attendance conditioned on an individual history all pay about the same
 contribution to the period-$N_c$ dynamics, resulting in a stronger
 correlation. However, the strength of this auto-correlation conditioned on a
 particular history does not give complete information on the degree of
 overcrowding. It is the disappearance of period-$N_c^{M+1}$ dynamics and the
 absence of virtual score reset mechanism that make a player more likely to
 stick to a strategy. Hence, players cooperate slightly better leading to a
 slightly smaller mean variance of attendance in other
 MG$_\textrm{\scriptsize sub}(N_c)$ \cite{LatestMemory}. In this way,
 crowd-anticrowd theory explains not only why all the $\Sigma^2 / N$ \emph{vs.}
 $\alpha$ curves in Fig.~\ref{F:Var} follows the same trend in the symmetric
 phase, but also attributes their slight differences to the strength of
 period-$N_c$ dynamics.

 We move on to discuss the situation of $\alpha \gtrsim \alpha_c$, namely, the
 asymmetric phase. In this phase, the number of strategies at play $N S$ is
 less than the maximal reduced strategy space size $N_c^{M+1}$. Thus, the
 probability that a particular alternative can never be picked by at least $N /
 N_c$ players increases as the number of players $N$ decreases. By pigeonhole
 principle, for an alternative to be the majority choice, the number of players
 choosing that alternative must be at least $N / N_c$. Consequently, whenever
 $\alpha \gtrsim \alpha_c$, some alternatives may never have a chance to be the
 majority choice. In contrast, there is no such type of constraint preventing
 an alternative from being a non-$\textrm{maj} (1)$ choice. Therefore, among
 all the history string generation methods we have studied, the $\textrm{sub} =
 \textrm{maj} (1)$ one will generate the most non-uniformly distributed history
 strings. This assertion is confirmed in Fig.~\ref{F:Freq_Dist}, which plots
 the frequency count of the history occurrence arranged in descending order and
 then averaged over 50 runs. As the non-uniformity of history string leads to a
 reduction of the effective strategy space size, the crowd-anticrowd
 cancellation is strengthened in the asymmetric phase \cite{Memory}. Thus, the
 mean variance of attendance obtained by using a majority history string is
 consistently lower than those obtained by using other history string
 generation methods. Again, crowd-anticrowd theory is able to explain the
 slight difference in $\Sigma^2 / N$ between $\textrm{maj} (1)$ and
 non-$\textrm{maj} (1)$ history posting methods in the asymmetric phase.

 Finally, we study the phase transition point $\alpha = \alpha_c$. The
 crowd-anticrowd analysis reported earlier attributes the increase in variance
 per player by decreasing $\alpha$ in the symmetric phase (increasing $\alpha$
 in the asymmetric phase) to over-crowding of strategies (insufficient
 sampling). In this respect, crowd-anticrowd theory predicts the existence of a
 single phase transition point separating the regions $\theta = 0$ and $\theta
 > 0$ around $\alpha \approx 1$ although it does not provide an efficient way
 to compute its value $\alpha_c$. Near the point of maximal cooperation, the
 number of strategy at play is approximately equal to $N_c^{M+1}$. In this
 regime, the number of players using strategies $(\lambda, \beta)$ and
 $(\lambda,\beta')$ are always about the same for all $\beta\neq \beta'$. Thus,
 from Eq.~(\ref{E:Sigma_CAC}), the small variance at the point of maximal
 cooperation is the result of an optimal crowd-anticrowd cancellation
 \cite{Crowd1,Crowd2,NcMG}. Recall that two anti-correlated strategies always
 give different suggestions irrespective of the history string. So, it seems
 reasonable that near the point of maximal cooperation, the degree of
 cooperation and hence the values of $\alpha_c$ and $\left. \Sigma^2 / N
 \right|_{\alpha_c}$ are independent of the history string generation method
 ``$\textrm{sub}$''. However, one should not regard this argument as a firm
 proof for the precise values of $\alpha_c$ and $\left. \Sigma^2 / N
 \right|_{\alpha_c}$ may depend in general on the complex adaptive dynamics of
 the system.

\section{Discussions}
 We have introduced a modification of the MG called
 MG$_\textrm{\scriptsize sub}(N_c)$ and studied its properties numerically. We
 argued under the framework of crowd-anticrowd theory that the general trend of
 the $\Sigma^2 / N$ \emph{vs.} $\alpha$ curve is independent of the real
 history string generation method although the dynamics of the system depends
 on the kind of common history string used. We also numerically find that
 $\alpha_c$ and $\left. \Sigma^2 / N \right|_{\alpha_c}$ are functions of $N_c$
 and $S$ only and are independent of the type of global history ``sub'' used.
 This finding is consistent with crowd-anticrowd theory. We remark that
 although all numerical simulations reported in the paper are performed by
 picking the strategies from the full strategy space, the same conclusions are
 reached if the strategies are taken from the maximal reduced strategy space.
 To summarize, these findings together with that in Ref.~\cite{LatestMemory}
 show that the level of cooperation among players does not change significantly
 if the minority history string is replaced by a variety of global data such as
 the historical majority and a fake history string.

 It is instructive to further extend our analysis to the case that the rankings
 of all the $N_c$ alternatives in each of the previous $M$ turns are used as
 public information. We believe that both kinds of periodic dynamics should be
 present in the symmetric phase. Verification of these hypothesis by numerical
 simulation is, however, a very computational intensive task as the maximal
 reduced strategy space size is prohibitively large except for $N_c \lesssim
 4$.

 Lastly, our findings imply that one must alter the original MG in some other
 ways in order to significantly change the trend of the $\Sigma^2 / N$
 \emph{vs.} $\alpha$ curve. Possibilities include the use of thermal updating
 rules \cite{Thermal} and the introduction of initial bias in virtual score
 \cite{Wong1,Wong2}.

\begin{acknowledgments}
 We would like to thank HKU Computer Center for their helpful support in
 providing the use of the HPCPOWER~System for most of the simulations reported
 in this paper. Useful discussions with C.~C. Leung is gratefully acknowledged.
\end{acknowledgments}

\bibliography{mg7.3}
\end{document}